\documentclass[aps,prl,twocolumn,showpacs,superscriptaddress]{revtex4-1}

\usepackage{amsmath,braket,amssymb,natbib,graphicx,float,amsthm,xcolor,outlines,mathrsfs,multirow,hhline}

\usepackage{dcolumn}% Align table columns on decimal point
\usepackage[english]{babel}
\usepackage[utf8x]{inputenc}
\usepackage[T1]{fontenc}
\usepackage{listings}
\usepackage[colorinlistoftodos]{todonotes} 
\usepackage[colorlinks=true, allcolors=blue]{hyperref}
\usepackage{enumitem}
\DeclareMathOperator{\nb}{\bar{{\it n}}}

\begin{document}
\title{Sub-Planck Structure Quantification in Non-Gaussian Probability Densities}
\author{Darren W. Moore}
\email{darren.moore@upol.cz}
\affiliation{Department of Optics, Palack\'{y} University, 17. listopadu 1192/12, 779 00 Olomouc, Czech Republic}
\author{Vojt\v{e}ch \v{S}varc}
%\email{}
\affiliation{Department of Optics, Palack\'{y} University, 17. listopadu 1192/12, 779 00 Olomouc, Czech Republic}
\author{Kratveer Singh}
%\email{}
\affiliation{Department of Optics, Palack\'{y} University, 17. listopadu 1192/12, 779 00 Olomouc, Czech Republic}
\author{Artem Kovalenko}
%\email{}
\affiliation{Department of Optics, Palack\'{y} University, 17. listopadu 1192/12, 779 00 Olomouc, Czech Republic}
\author{Minh Tuan Pham}
%\email{}
\affiliation{Institute of Scientific Instruments of the Czech Academy of Sciences, Kr\'{a}lovopolsk\'{a} 147, 612 64 Brno, Czech Republic}
\author{Ond\v{r}ej \v{C}\'{i}p}
%\email{}
\affiliation{Institute of Scientific Instruments of the Czech Academy of Sciences, Kr\'{a}lovopolsk\'{a} 147, 612 64 Brno, Czech Republic}
\author{Lukáš Slodička}
%\email{slodicka@optics.upol.cz}
\affiliation{Department of Optics, Palack\'{y} University, 17. listopadu 1192/12, 779 00 Olomouc, Czech Republic}
\author{Radim Filip}
%\email{filip@optics.upol.cz}
\affiliation{Department of Optics, Palack\'{y} University, 17. listopadu 1192/12, 779 00 Olomouc, Czech Republic}

\begin{abstract}
Sub-Planck structures in non-Gaussian probability densities of phase space variables are pervasive in bosonic quantum systems. They are almost universally present if the bosonic system evolves via nonlinear dynamics or nonlinear measurements. So far, identification and comparison of such structures remains qualitative. Here we provide a universally applicable and experimentally friendly method to identify, quantify and compare sub-Planck structures from directly measurable or estimated probability densities of single phase space variables. We demonstrate the efficacy of this method on experimental high order Fock states of a single-atom mechanical oscillator, showing provably finer sub-Planck structures as the Fock occupation increases despite the accompanying uncertainty increase in the phonon, position, and momentum bases.
\end{abstract}

\maketitle

{\it Introduction}---~The uncertainty principle in quantum mechanics limits the simultaneous precision of non-commuting measurements to $\Delta x\Delta p\ge\frac{\hbar}{2}$. This defines a minimum area scale for a quantum state in the Wigner function representation of phase space based on the canonical variables $x$ and $p$, proportional to $\hbar$. The phrase `sub-Planck structure' then refers to qualitative features of a phase space distribution that are confined to regions with an area much smaller than $\hbar$~\cite{zurek_sub-planck_2001} i.e. smaller than the ground state extension. It remains open whether these features can also provide uncertainties below this extension. Nevertheless it has been recognised that not only are such structures physically meaningful, but that they are highly relevant for the sensitivity of the quantum state, whether to an environment or as a probe, and more generally for the noise reduction required for improving quantum technologies. Area scales in phase space are also relevant for classical Hamiltonian and quantum mechanics due to symplectic geometry~\cite{hsiao_fundamental_2007,de_gosson_symplectic_2009,de_gosson_symplectic_2013} but the uncertainty principle persists beyond the trace-preserving (deterministic) symplectic transformations of unitary Gaussian quantum mechanics~\cite{dragt_how_1998,weedbrook_gaussian_2012} into all quantum processes. Consequently area scales below those of the uncertainty principle require a more sophisticated yet operational strategy to interpret them in terms of familiar and physical nonclassical features.

These sub-Planck structures are qualitatively visible in the (possibly negative) phase space sub-structure of non-Gaussian quantum states even if the variances of the phase space variables are above those of the ground state. Unfortunately, the Wigner function's status as a quasi-probability distribution, particularly the negative values, does not allow for the definition of sub-Planck structure using the spread in phase space. This problem has been tackled unsatisfactorily many times since the pioneering kitten state reconstructions~\cite{meekhof_generation_1996,leibfried_experimental_1997,ourjoumtsev_generating_2006,deleglise_reconstruction_2008,hacker_deterministic_2019,magro_deterministic_2023} and recently revived for large cat states~\cite{deleglise_reconstruction_2008,vlastakis_deterministically_2013,johnson_ultrafast_2017,lewenstein_generation_2021,kudra_robust_2022,bild_schrodinger_2023,pan_protecting_2023}, binomial states~\cite{hu_quantum_2019,kudra_robust_2022}, Gottesman-Kitaev-Preskill states~\cite{fluhmann_encoding_2019,campagne-ibarcq_quantum_2020}, multi-squeezed states~\cite{eriksson_universal_2024}, cubic phase states~\cite{kudra_robust_2022,eriksson_universal_2024}, and many other non-Gaussian states~\cite{yukawa_generating_2013,shalibo_direct_2013,kirchmair_observation_2013,he_fast_2023,marti_quantum_2024,jeng_strong_2024}. For example, an effective squeezing introduced for the Gottesman-Kitaev-Preskill states has been used to specify its quality for error correction~\cite{duivenvoorden_single-mode_2017,takase_generation_2024} but this does not generalise.

Therefore it remains an outstanding problem to identify, quantify and compare sub-Planck structures across the exceptionally diverse forms in which they may emerge. Even without negative phase space values, mixtures of nonclassical states may feel classical non-Gaussian noise, and even more surprisingly, quantum non-Gaussian states may be positive even outside the convex completion of the Gaussian states~\cite{filip_detecting_2011}, obscuring the sub-Planck structure. Currently, the typical approaches to dealing with quantum states possessing complex phase space structures is to look to a kind of hierarchical taxonomy of quantum states, with the simplest involving notions of nonclassicality and quantum non-Gaussianity~\cite{lachman_quantum_2022,rakhubovsky_quantum_2024}, and more complex classifications into Fock state hierarchies~\cite{lachman_faithful_2019} or finite stellar rank~\cite{chabaud_stellar_2020}. The problem is instead solved by providing an operational and reliable quantification that can conclusively identify, quantify and compare sub-Planck structures in any nonclassical state, including those with infinite stellar rank.

\begin{figure*}
    \centering
    \includegraphics[width=2\columnwidth]{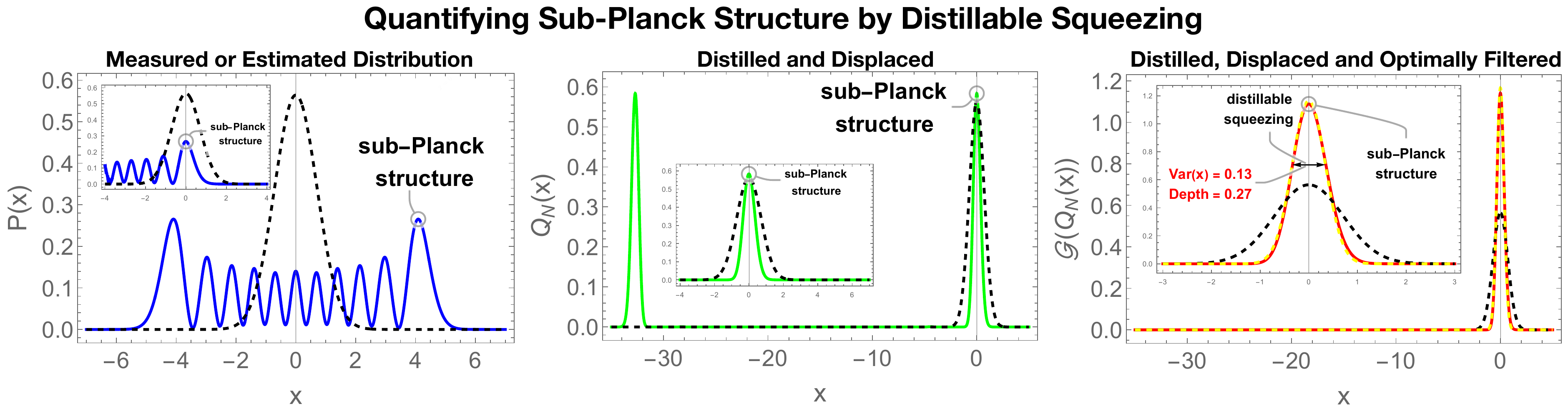}
    \caption{Schematic of the computational procedure quantifying the nonclassical sub-Planck structure around the global maximum of the continuous variable distribution, whose distillable squeezing asymptotically approaches that given by Eq.~(\ref{Asym}). The quantification of sub-Planck structure requires the estimation of the probability distribution $P(x)$ to be distilled, followed by an operationally defined processing of the data. This processing produces an output distribution $Q_N(x)$ with a global maximum displaced to the origin followed by optimised filtration $\mathcal{G}\left(Q_N(x)\right)$. This is illustrated for the discrete variable experimental Fock state $\ket{10}$. The black dashed lines are the ground state and the insets show a direct comparison of the global maximum with the ground state variance. The sub-Planck structure embedded in the relative concavity of $P(x)$ around a global maximum is preserved by the distillation while being promoted from a local feature to a global feature: the variance of the distilled distribution $\mathcal{G}\left(Q_N(x)\right)$. The dashed yellow line shows the Gaussian squeezed state whose variance is calculated from Eq.~(\ref{Asym}).}
    \label{Sketch}
\end{figure*}

To this end we recontextualise and advance the method of experimental distillation of quantum squeezing~\cite{heersink_distillation_2006,franzen_experimental_2006,marek_direct_2017}, into a theoretical certification method based on~\cite{filip_distillation_2013} which universally and conclusively identifies nonclassical sub-Planck structures in phase space, and uncovers and compares new nonclassical aspects. Distillable squeezing detects sub-Planck structure manifested as nonclassical quantum squeezing in a phase space variable, even for non-Gaussian states having no squeezing~\cite{filip_squeezed-state_2014}. The detection is successful even in the absence of negative values of the Wigner function or phase sensitive structures~\cite{marek_multiple-copy_2007}. We emphasise that we are methodologically reframing the experimental distillation of squeezing to a theoretical quantifier of sub-Planck structures. Therefore our approach advantageously converts a less efficient (conditional) state preparation process into a versatile identification method as the distillation takes place virtually in a classical computer. That is, although the distillation is conditional we do not suffer from the exponentially decreasing probability of success typical of state preparation protocols. The result is a method which provides a conclusive, operational and deterministic recognition of nonclassical sub-Planck structure in which all single mode quantum states are commensurable.

{\it Sub-Planck Structure by Distillable Squeezing}---~Essentially, the squeezing that could {\it in principle} be distilled from an arbitrary number of copies of a given state can be given conceptual primacy. Importantly, this shift in perspective, unlike the preparation of squeezed states by distillation~\cite{heersink_distillation_2006,franzen_experimental_2006,marek_direct_2017}, does not require physically carrying out the distillation, provided that the evaluation of distillable squeezing is clearly and operationally defined. It only requires knowledge of the measured or deduced probability distribution of the phase space variable whose distillable squeezing can be deterministically computed and which can be varied to investigate other sub-Planck structures. The distillable squeezing from this distribution is then used as a conclusive measure of the sub-Planck structure and renders otherwise incomparable non-Gaussian states numerically commensurable.

\begin{figure*}
    \centering
    \includegraphics[width=\linewidth]{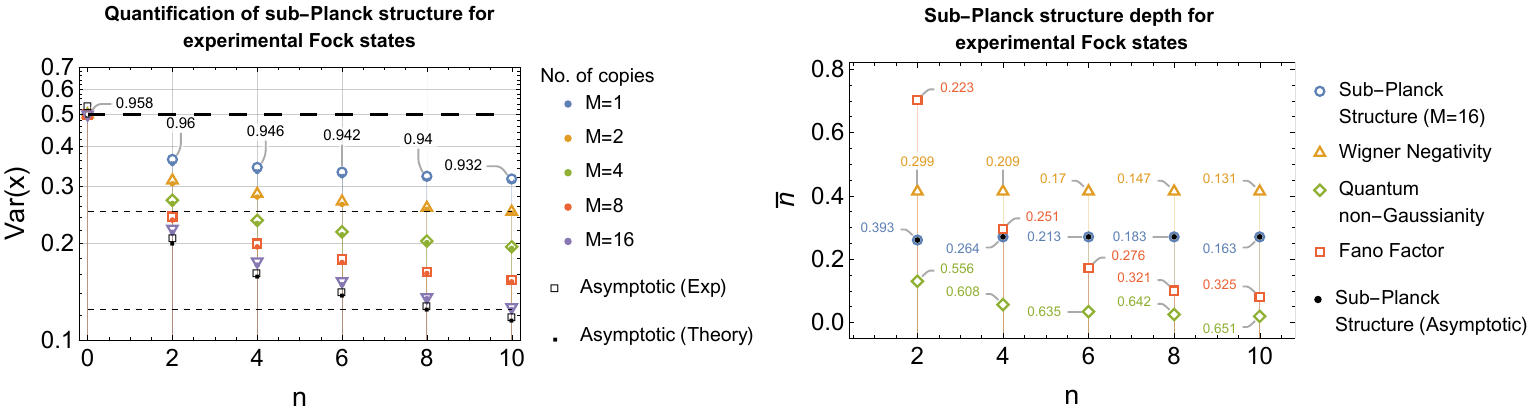}
    \caption{Left: Sub-Planck structure quantification by universal distillation of squeezing from experimental Fock states (empty markers), compared with the pure states (solid markers), as a function of the occupation $n$, where $n\in\{0,2,4,6,8,10\}$. The heavy dashed line indicates the ground state noise, while the upper and lower dashed lines represent 3~dB and 6~dB of squeezing respectively. The numbers above points are the probabilities associated with the experimental Fock state $\ket{n}$. Larger $n$ results in consistently larger distillable squeezing, confirming that larger Fock states possess greater nonclassical sub-Planck structure. The colours indicate the number of copies $M=2^N$ used in the distillation process. Increasing the copies consumed consistently results in an increase in distillable squeezing. Black squares indicate the asymptotic limit of distillable squeezing [Eq.~(\ref{Asym})] for the experimental (empty squares) and pure (solid squares) Fock states.
    Right: The sub-Planck structure depth for the experimental Fock states under thermalisation, with $M=16$, and compared with the thermalisation depth of Wigner negativity, the thermalisation depth for $n$-photon QNG criteria calculated from the Fock state distribution, and the thermalisation depth for sub-Poissonian statistics (Fano factor). The coloured numbers connected to the points are the probabilities for the respective Fock state after thermalisation. For the pure states the sub-Planck structure depth is consistently $\nb\approx0.28$, compared with the Wigner negativity which vanishes at $\nb=0.5$ for all $n$, and the $n$-photon QNG depth which strictly decreases with $n$ (see SM). For the experimental states, composed of a finite mixture of Fock states, the sub-Planck structure depth advantageously remains constant and similar to that of the pure states.}
    \label{FockDist}
\end{figure*}

In order for the distillable squeezing to simultaneously provide some quantification of sub-Planck structure and satisfy the logic of commensurability the method of distillation must be fixed, i.e. universal, rather than adapted to each input state. That is, it is not relevant for us to optimise the distillation procedure to maximise the distillable squeezing; instead the method is specified ahead of time and used universally across all detected distributions (see Fig.~\ref{Sketch}). We refer to the selected method as universal quantification of the sub-Planck structure. The quantification must be faithful, therefore the distillation method cannot produce any nonclassicality from classical states. This sharply constrains the possible procedures. The method we will choose as universal will consist only of linear interference, phase space variable measurements, linear displacements, and an ancillary harmonic oscillator ground state. This universal and operational method identifies the sub-Planck structure present around the global maximum $a$ of the measured distribution. This manifests as the local relative concavity around this maximum (see Fig.~\ref{Sketch}), which provides an asymptotic value~\cite{filip_distillation_2013} of the distillable squeezing as
\begin{equation}\label{Asym}
    \text{Var}(x)=\frac{P(a)}{|P^{\prime\prime}(a)|}\,,
\end{equation}
in the limit of a sufficiently large number of copies of the distribution $P(x)$, where now we let $x=q\cos\phi+p\sin\phi$ be an arbitrary phase space variable. The local relative concavity around this global maximum is preserved by the distillation (see Fig.~\ref{Sketch}, and SM) and promoted to a global feature approaching a Gaussian distribution. That is, $\frac{Q^{\prime\prime}(a)}{Q(a)}=\frac{P^{\prime\prime}(a)}{P(a)}$ where $Q(x)$ is the output distribution of the distillation~\cite{filip_distillation_2013}.

The operational meaning behind formula (\ref{Asym}) is the use of multiple copies of the measured distribution $P(x)$ of the selected phase space variable. To reach the asymptotic formula (\ref{Asym}), the distributions are then processed analytically or in a computer mimicking linear symmetrical interference of multiple distributions with conditioning and postselection on $x$. The output of this data processing may be sent through a further filter based on both linear splitting and detection to single out only one global maximum. None of these steps can create nonclassical states from classical states, therefore the quantification given by (\ref{Asym}) is faithful. The variance of the resulting distribution determines the distillable squeezing and therefore the presence of nonclassical sub-Planck structure in that phase space variable. 

Two copies are sufficient to illustrate. Two identical copies of the estimated distribution, $P(x_1)$ and $P(x_2)$, experience linear interference such that the output joint distribution is $P\left(\frac{x_1+x_2}{\sqrt{2}}\right)P\left(\frac{x_1-x_2}{\sqrt{2}}\right)$. One variable is then conditioned over the measurement result $\bar{x}=0$, leaving the (unnormalised) distribution $P\left(\frac{x}{\sqrt{2}}\right)^2$ (see SM). The distillation procedure over $N$ layers with $M=2^N$ copies results in the output density function $Q_N(x)\propto P\left(\frac{x}{\sqrt{M}}\right)^{M}$.
% \begin{equation}\label{distill}
%        Q_N(x)\propto P\left(\frac{x}{\sqrt{M}}\right)^{M}\,.
% \end{equation} 
In the case in which there are multiple global maxima, these are preserved by the distillation, along with their relative concavity and become increasingly separate from each other as $N$ increases, artificially inflating the variance and obscuring the sub-Planck structure quantification. This is overcome by an additional data filtration step, computationally simulating the experimental procedure~\cite{heersink_distillation_2006}, which filters out the distribution around a chosen global maximum while preserving the concavity. This filtration, which also cannot create nonclassical states from classical states, consists of first displacing the distilled state $Q_N(x)$ so that one global maximum lies at the origin. This single copy is filtered by linear interference with the oscillator ground state followed by a conditioning step again on $\bar{x}=0$. A successful filtration simply corresponds to maximising the distillable squeezing by optimising the linear interference. The limiting distillable squeezing obtained from this procedure is exactly that of Eq.~(\ref{Asym}).

{\it Sub-Planck Structures in Experimental Fock States}---~To test this methodology we first apply it to the sub-Planck structures in high Fock states, that is we test the procedure in detecting continuous-variable nonclassical squeezing in the  complementary basis of discrete variable states. To test the quantification and compare our approach we adopt the experimental setup and procedures used to prepare high quality and provably quantum non-Gaussian Fock states in the motion of a single trapped ion~\cite{lachman_faithful_2019}. We generated new data to quantify the sub-Planck structures, see details in Supplementary Material (SM). The Fock states have rotational symmetry so that there is no need to decide from which quadrature to distill squeezing. In other words, we can separately detect the same sub-Planck structure in position and momentum below the ground state wavepacket extension. Moreover the deduced quadrature distributions used in the distillation always possess multiple global maxima which allows us to illustrate how to deal with this problem directly through auxiliary filtration.

The Fock states $\ket{n}$ have quadrature probability distributions which are described by Gaussian modulated Hermite polynomials~\cite{ferraro_gaussian_2005}. Here we choose units such that the ground state variance is $\frac12$, identical for both position and momentum. In Fig.~\ref{FockDist} we show the distillable squeezing for various numbers of copies $M$ of the input Fock states. As the Fock state occupation increases, the amount of distillable squeezing from sub-Planck structure also increases. More significantly, as the number of copies increases, the distillable squeezing approaches the asymptotic value and the rate of increase in distillable squeezing also increases at least up to $\ket{10}$ and $M=16$, as can be seen in the broadening of the distribution of data points going left to right. In fact the efficiency of the distillation increases with the number of copies (see SM). Despite the residual impurity (phonon noise) of the experimental Fock states the distillable squeezing remains robust and similar to that of the theoretical pure states, in both the finite copy and asymptotic cases. This robustness remains although both quadrature and phonon number variances increase. This is a significant test of the robustness of the sub-Planck quantifier, as it demonstrates that it can be applied to non-Gaussian states of realistic quality. This is further supported by the inclusion of the imperfect ground state as it is a classical Gaussian state as well as a Fock state. With added phonon noise, sub-Planck structure can artificially be distilled from the errors, which are all contributions from nonclassical Fock states. Yet the distillable squeezing from such noise remains negligible. The right panel shows the robustness of the distillable squeezing to thermalisation, which we refer to as sub-Planck structure depth (see SM for definitions and an extended discussion).

The sub-Planck structure for the pure Fock states is sensitive to thermalisation, with a typical depth of $\nb\approx0.28$, which may also be calculated directly from Eq.~(\ref{Asym}) applied to theoretically thermalised Fock states (see SM). This may be further compared with $\nb=\frac12$ for the pure state Wigner negativity depth. The sub-Planck structures of Fock states are therefore more sensitive than the negative points of the Wigner function under thermalisation. Moreover their depth does not appreciably change with the order of the Fock state, in contrast to that of their quantum non-Gaussian hierarchy~\cite{podhora_quantum_2022} or their sub-Poissonian statistics as measured by the Fano factor~\cite{fano_ionization_1947}. This reduces the demand for much smaller error bars which would substantially prolong the measurements required to reach the same confidence for the higher Fock states. This positive stability is an indication that the sub-Planck structure of the high Fock states might be a stable resource and a reasonable quantifier for other quantum non-Gaussian states. A further stability is evinced by the effect of experimental phonon noise on the depth, in the panel on the right. The sub-Planck structure depth is stable under increasing $n$ with experimental fluctuations in the phonon number similar to the stability of the Wigner negativity depth (see SM for a further comparison). Furthermore the noise in the phonon distribution has little effect on the susceptibility of the distillable squeezing to thermalisation as compared with the pure states, as simulated thermalisation of the experimental Fock states closely matches that of the pure states (see SM). It can also be seen from the blue numbers in the right panel of Fig.~\ref{FockDist} that nonclassical sub-Planck structure can still be detected even when the probability of the relevant Fock state is greatly reduced, even below $\frac12$.

{\it Conclusion and Outlook}---~We have provided a universal methodology for the quantification of sub-Planck structure and its depth via distillable squeezing connected to the relative concavity of the maximum of a non-Gaussian probability density. This methodology can be more widely applied to states in which non-Gaussian noise or Wigner positivity hides the sub-Planck structure. With only one selected quantum phase space variable it is possible to conclusively and operationally quantify sub-Planck structure from general non-Gaussian and even phase insensitive states with sufficient sub-Planck structure depth. Use of the phase space variable is advantageous, as distillation in the photon number basis is not efficient~\cite{berry_linear-optical_2010}. The method has been tested with experimentally obtained Fock states which climb the genuine QNG (stellar) hierarchy. An extremely high order Fock state of twenty phonons was also tested and is stable against phonon noise with distillable squeezing approaching the ideal asymptotic value. As an example of more general applicability, we apply the universal method successfully to ideal cat states and GKP states (see SM). The universal distillable squeezing and its depth can then be compared with the ordering set by the hierarchical Fock states, even for states with inconclusive stellar rank. Moreover, the approach can be straightforwardly extended to multimode distillable squeezing (or distillable generalised squeezing) to quantitatively study sub-Planck structures in the bosonic correlatons of phase space variables.

Additionally the phase space interference phenomena that give rise to sub-Planck structures are not confined to the concavity around a global maximum and a {\it non-universal} approach evading this confinement will unveil even more diverse features of nonclassical sub-Planck structures even in cases where no such structure can be detected at the global maximum. As a critical example, the momentum distribution of the cubic phase state cannot be universally distilled. This limitation can be overcome by a single preparatory non-universal distillation layer where the conditioning step in the distillation is not at $\bar{x}=0$ (compare with distillation procedure in SM). This step also cannot prepare nonclassical states from classical ones, but it can prepare non-Gaussian densities with narrow concavities, which are then susceptible to the universal method described here. Thus the nonclassical sub-Planck structures of states with inconclusive stellar rank can be dealt with using the method of distillable squeezing. 

\begin{acknowledgments}
The authors acknowledge the project 22-27431S of the Czech Science Foundation, project CZ.02.01.01/00/22\_008/0004649 (QUEENTEC) and 8C22001 (SPARQL) of MEYS Czech Republic and the EU. This work
was also jointly supported by the SPARQL and CLUSTEC projects that have received funding from the European Union's Horizon 2020 Research and Innovation Programme under Grant Agreement No. 731473 and 101017733 (QuantERA) and No. 101080173 (CLUSTEC).
\end{acknowledgments}

\bibliographystyle{unsrt}
\bibliography{references}

\appendix

\onecolumngrid

\section{Experimental Fock State Generation and Number Estimation}

We generate the experimental Fock states on the motion of a single $^{40}\mathrm{Ca}^+$ ion in a linear Paul trap. Applying a 29.9-MHz radio-frequency electric potential of 1250~V (peak-to-peak) to a pair of radial electrodes and static electric potentials of about 1000~V at the two axial tip electrodes results in a secular axial motional frequency of 1.1 MHz and approximately 2 MHz for the axial and near-degenerate radial motions of the ion. A magnetic field of 3.3~Gauss is applied along the trap axis to lift the degeneracy of Zeeman states. Interrogation of the  $4^2\textrm{S}_{1/2} \leftrightarrow 3^2\textrm{D}_{5/2}$ transition is performed using a frequency-stabilized diode laser at 729 nm, with the laser beam directed at a 45-degree angle relative to the trap axis. This configuration produces a Lamb-Dicke parameter of $\eta=0.063$ and a carrier Rabi frequency of $\left( 2 \pi \right)  \times$~32~kHz.

The process of creating a mechanical Fock state begins by preparing the system in the motional ground state. First, Doppler cooling is performed using a red-detuned 397~nm laser on the $4^2 \textrm{S}_{1/2} \leftrightarrow 4^2 \textrm{P}_{1/2}$ transition, along with an 866~nm laser to reshuffle the population of $3^2 \textrm{D}_{3/2}$ manifold back to the cooling transition. Sideband cooling is realized on the $4^2\textrm{S}_{1/2} (m=-1/2) \leftrightarrow 3^2\textrm{D}_{5/2}, (m=-5/2)$ transition along with short pulses of \(\sigma\)-polarized light 397~nm for optical pumping and an 854~nm laser for rapid reshuffling to the \(\left|4^2\textrm{S}_{1/2}, m = -1/2\right\rangle\) state. Next, a sequence of resonant blue-sideband and red-sideband \(\pi\)-pulses on \( 4^2\textrm{S}_{1/2} (m=-1/2)  \leftrightarrow  3^2\textrm{D}_{5/2} (m=-1/2) \) is implemented to coherently transfer the population from the motional ground state to the target Fock state. The resulting motional populations are then measured by observing Rabi oscillations on the blue motional transition. This measurement is repeated 100~times to minimize the effects of quantum projection noise on the estimated excited state probability. Finally, the Fock state occupancy \(P_n\) is reconstructed by fitting the accumulated data sets with
\[
P_e = \sum_{n=0}^{\textrm{N}} P_n \sin^2 \left( \frac{\Omega_{n,n+1} \, t}{2} \right)
\exp \left[-\gamma  t (n+1)^{0.7}\right]
\]
where $P_{\rm e}$ is the probability of projection to the excited state, $\Omega_{n,n+1}$ represents Rabi frequency of interaction including motional Fock states $\ket{n}$, $\ket{n+1}$, and $\gamma$ quantifies a decay~\cite{leibfried_quantum_2003}. To determine the $\Omega_{n,n+1}$, we use the scaling outside the Lamb-Dicke regime to avoid a mismatch between the fitted and real Rabi frequency. Specifically, having a Lamb-Dicke factor of~0.063, the Rabi frequency mismatch for $\Omega_{10,11}$ is expected to be about 2~\%.
Since $\Omega_{n,n+1}$ scales approximately with a square root of~$n$, Rabi frequencies between neighboring Fock states become similar $\Omega_{n,n+1}\approx\Omega_{n+1,n+2}$ for higher~$n$. As an example, for $\ket{10}$ and $\ket{11}$ the ratio $\frac{\Omega_{10,11}}{\Omega_{11,12}}\approx 0.96$. Consequently, the reliable estimation of higher Focks becomes nontrivial. To make the reconstruction robust, first, the Rabi flop scanning time and data sampling are set accordingly to the Fock state number. Higher Fock states require denser sampling and more Rabi flops scanned. Also, we characterize precisely the decay factor $\gamma$ and ground state Rabi frequency $\Omega_{0,1}$, and use those values in the data reconstruction process to get more accurate results. 

The observed populations of the desired Fock states are mainly limited by the following phenomena. Imperfect initialization to the motional ground state, which corresponds to a thermal state with a mean phonon number of approximately~0.05, can limit the population of $\ket{10}$ to below~95.2\%. This can be reduced through more optimal ground-state cooling (a mean phonon number of approximately 0.02 has been achieved) or by performing a ground-state postselection before Fock state preparation, both at the significant cost of increased measurement duration. During the state preparation sequence, motional heating causes thermal redistribution of population from the actual Fock states $\ket{n}$ to $\ket{n-1}$ and $\ket{n+1}$. With the measured motional heating rate of 3~phonons per second at the employed motional frequency, the total duration of the Fock state generation sequence of 1.3~ms for $\ket{10}$ sets a limit on the $P_{10} \leq 96.4\%$ due to the corresponding tstate thermalization effects. This can be further improved by adjusting the Lamb-Dicke parameter or using higher 729~nm laser power to speed up the Fock preparation sequence. The imperfect population transfer from the motional ground state to the desired Fock state is also constrained by the finite coherence of laser-ion interactions, estimated to around $10$~ms for the carrier transition, and the motional coherence of about $80\pm 4$~ms measured for the $\ket{0}+\ket{1}$ superposition states~\cite{turchette_decoherence_2000,kovalenko_quantum_2025}. Additionally, residual off-resonant excitations of the carrier transition, estimated at around 0.3~\% per pulse, contribute to the infidelity of each Rabi pulse on motional transitions. Together, these effects restrict $P_{10}$ to a range of approximately 88-91\%, which aligns with the experimental data within the error margins. The actual probability may vary slightly depending on the specific experimental parameters.

\section{Simulated Thermalisation}

To more deeply understand the effect of a thermalising environment on the distillable squeezing, we use the concept of {\it depth}, which measures the amount of thermalisation of the state required to remove the distillable squeezing. The thermalisation map on a state $\rho$ is given by 
\begin{equation}
M_{\nb}(\rho)=\int\frac{d^2\alpha}{\pi\nb}e^{-\frac{|\alpha|^2}{\nb}}D(\alpha)\rho D^\dagger(\alpha)\,,
\end{equation}
where $\nb$ is an effective temperature and $D(\alpha)$ is the displacement operator. This models a collection of phase randomised displacements in phase space, with the probability of the amplitude of the displacements scaled by $\nb$. To illustrate, if $\rho$ is the ground state $\ket{0}\bra{0}$ then thermalisation produces the oscillator thermal state. 

The effect of the thermalisation on the canonical quadrature probability distributions can be calculated directly from this map. For the position quadrature, the thermalised probability distribution is
\begin{equation}
P_{\nb}(x)=\int\frac{d^2\alpha}{\pi\nb}e^{-\frac{|\alpha|^2}{\nb}}\bra{x}D(\alpha)\rho D^\dagger(\alpha)\ket{x}\,,
\end{equation}
where for pure states $\rho=\ket{\psi}\bra{\psi}$ the expression simplifies to
\begin{equation}
P_{\nb}(x)=\int\frac{d^2\alpha}{\pi\nb}e^{-\frac{|\alpha|^2}{\nb}}|\bra{x}D(\alpha)\ket{\psi}|^2\,.
\end{equation}

\begin{figure}
\centering
\includegraphics[width=0.49\columnwidth]{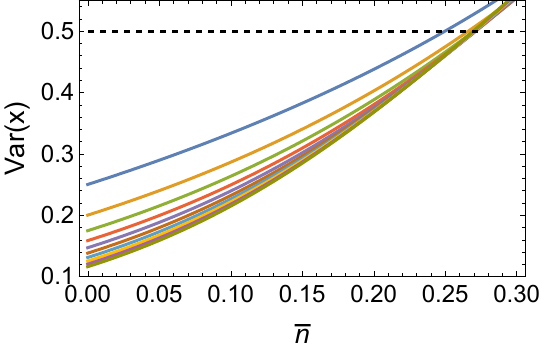}
\includegraphics[width=0.49\columnwidth]{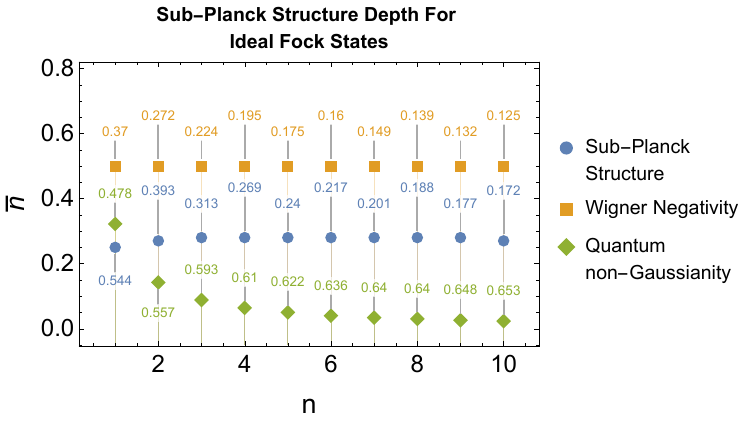}
\caption{Left: The sub-Planck structure depth for the asymptotic distillable squeezing calculated from the simulated thermalisation of the pure Fock states. The depth converges around $\nb\approx0.28$. Right: The sub-Planck structure depth compared with the QNG hierarchy depth and Wigner negativity depth for the pure Fock states.}
\label{AsymDepth}
\end{figure}

To analyse the depth of the distillable squeezing, we replace the input probability distribution $P(x)$ with the thermalised version $P_{\nb}(x)$. For Fock states, $\ket{\psi}=\ket{n}$, we have that $|\bra{x}D(\alpha)\ket{n}|^2=|\psi_n\left(x-\sqrt{2}\Re(\alpha)\right)|^2$, where $\psi_n(x)=\frac{e^{-\frac{x^2}{2}}}{\pi^{\frac14}}\frac{\text{H}(x)}{2^\frac{n}{2}\sqrt{n!}}$ are the eigenfunctions of the harmonic oscillator. In the case of the presence of phonon noise, $P_{\nb}(x)$ becomes a mixture distribution over the $\psi_n(x)$. In both cases the integral over $\alpha$ can be carried out exactly. The distillable squeezing is sensitive to thermalisation, possessing depth at around $\nb\approx0.28$ (see Fig.~\ref{AsymDepth}). This can be confirmed by testing the sub-Planck structure depth for the asymptotic values of the distillable squeezing which can be inferred using the formula in Eq.~(\ref{Asym}) and the thermalised Fock state quadrature distributions. The asymptotic distillable squeezing can then be calculated exactly as a function of $\nb$. For example, the thermalised Fock state $\ket{1}$ has the probability distribution
\begin{equation}
P_\text{th}^{(1)}=\frac{2e^{-\frac{x^2}{1+2\nb}}\left(x^2+2\nb^2+\nb\right)}{\sqrt{\pi}(1+2\nb)^\frac52}
\end{equation}
The relative concavity around the global maximum $x=\sqrt{1+\nb-2\nb^2}$ is easily calculated to be 
\begin{equation}
\frac{1+2\nb}{4|1-\nb|}\,,
\end{equation}
which only shows asymptotic distillable squeezing for $\nb\le\frac14$. The depth increases slightly beyond $n=1$ and stabilises at $\nb\approx0.28$, as numerically estimated in the main text for finite copies $M$. This is displayed in Fig.~\ref{AsymDepth}.

\section{Phonon Noise and Depth}

\begin{figure}
\centering
\includegraphics[width=\columnwidth]{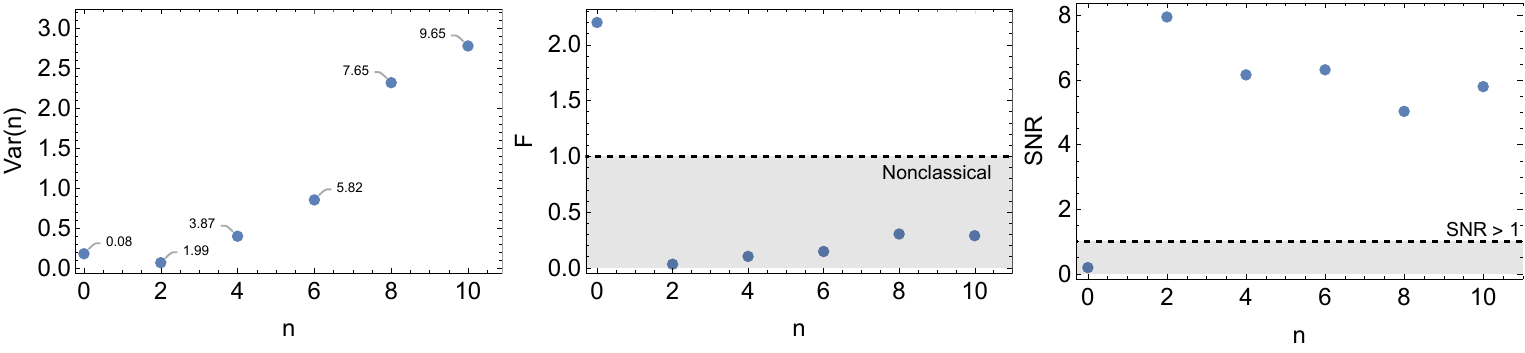}
\caption{Statistics for the experimental Fock states. Left: The variance increases for higher order Fock states, however the mean values (callout numbers) stay close to $n$. Center: The Fano factor $F=\frac{\text{Var}(n)}{\braket{n}}$, which is a detector of nonclassicality. $F<1$ for all experimental Fock states except the ground state. Right: The signal-to-noise ratio ($\text{SNR}=\frac{\braket{n}}{\sqrt{\text{Var}(n)}}$) of the experimental Fock states. The rising variance (left) does not outcompete the increase in mean value so that the SNR stays above 1, although it falls for higher $n$.}
\label{Fano}
\end{figure}

The experimentally generated states are characterised by a finite mixture of Fock states. That is, while these generated Fock states are of very high quality, sufficient to satisfy their genuine $n$-phonon quantum non-Gaussianity criteria, there remain small fluctuations in phonon number. This can be seen directly from the very high but non-unit probabilities in Fig.~\ref{FockDist} and also in the statistical analysis of Fig.~\ref{Fano}. This phonon noise affects both the distillable squeezing and the sub-Planck structure depth. Both properties are robust to phonon noise in the following ways.

\begin{figure}
\centering
\includegraphics[width=0.5\columnwidth]{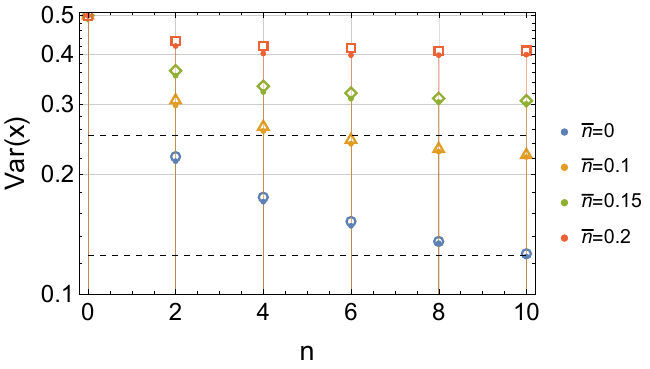}
\caption{The effect of phonon noise on the distillable squeezing for the experimental Fock states (empty markers) under thermalisation, with $M=16$ as compared with the theoretical results for the corresponding pure states (solid markers).}
\label{BothDepth}
\end{figure}

Recall that Fig.~\ref{FockDist} shows that the experimental and pure Fock states have similar distillable squeezing. In addition to this similarity, Fig.~\ref{BothDepth} shows that the experimental and pure Fock states respond similarly to thermalisation at the level of distillable squeezing. The sub-Planck structure is not made more sensitive to thermalisation via increases in the phonon noise.

\section{Operational Derivation of Distillable Squeezing}

\begin{figure}
\centering
\includegraphics[width=\columnwidth]{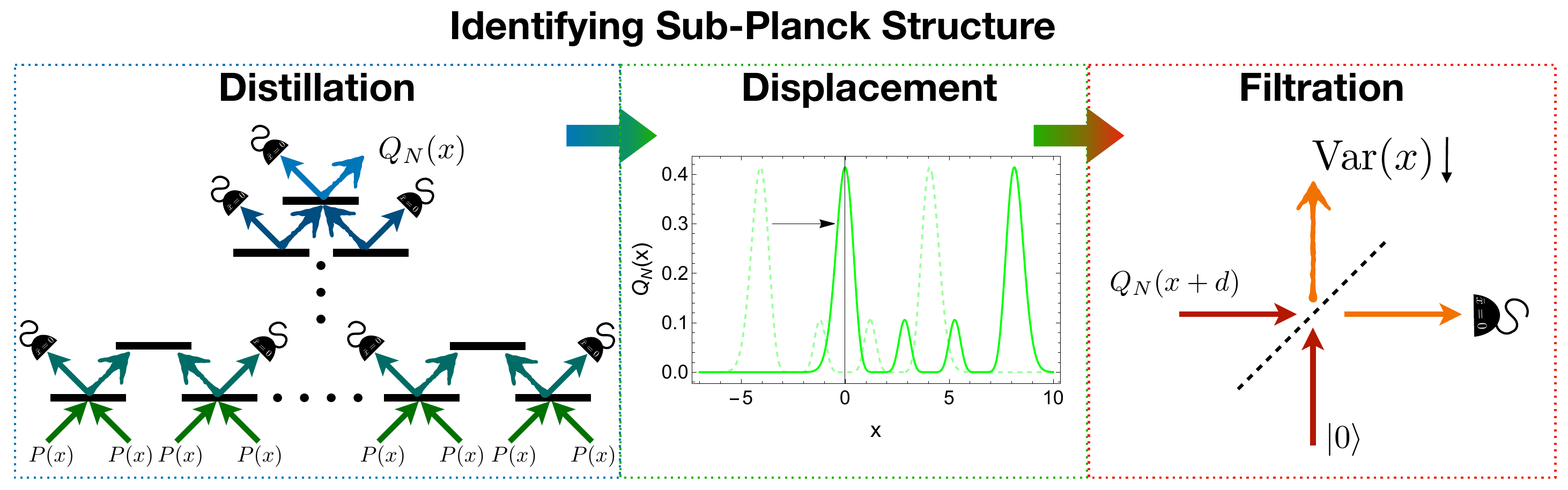}
\caption{Operational interpretation of Fig.~\ref{Sketch}. Each step executes quantum operations which do not produce nonclassicality.}
\label{Operational}
\end{figure}

The procedure to calculate the distillable squeezing of a quantum state depends only on the ability to accurately estimate the required probability distribution. The use of disillable squeezing as a detector of non-classical sub-Planck structure is itself built on a realisic distillation procedure that could in principle be carried out, albeit with vanishing probability of success. Here we show how the output distribution from the main text results from a linear optics inspired procedure provided in Fig~\ref{Operational}. 

For a pair of variables, $x_1$, $x_2$, both distributed according to $P(x)$ and split on a balanced beamsplitter, the output distribution takes the form $P\left(\frac{x_1+x_2}{\sqrt{2}}\right)P\left(\frac{x_1-x_2}{\sqrt{2}}\right)$. After the measurement, the conditional distribution takes the form
\begin{equation}
Q_1(x)=\frac{1}{\mathcal{N}}P\left(\frac{x_1+\bar{x}}{\sqrt{2}}\right)P\left(\frac{x_1-\bar{x}}{\sqrt{2}}\right)\,,
\end{equation}
where $\mathcal{N}$ is a normalisation. At a second layer, a pair of these conditional distributions is input to the beamsplitter and measured, resulting in
\begin{equation}
Q_2(x)=\frac{1}{N_1^2N_2}P\left(\frac{x+\bar{x}}{(\sqrt{2})^2}+\frac{\bar{x}}{\sqrt{2}}\right)P\left(\frac{x+\bar{x}}{(\sqrt{2})^2}-\frac{\bar{x}}{\sqrt{2}}\right) P\left(\frac{x-\bar{x}}{(\sqrt{2})^2}+\frac{\bar{x}}{\sqrt{2}}\right)P\left(\frac{x-\bar{x}}{(\sqrt{2})^2}-\frac{\bar{x}}{\sqrt{2}}\right)\,.
\end{equation}
where $\mathcal{N}$ is a new normalisation. 

To go to an arbitrary number of layers, let $c\in C^N=\{0,1\}^N$ be a member of the set of binary sequences of length $N$. At the final layer then, the distilled distribution is
\begin{equation}
Q_N(x)=\frac{1}{\mathcal{N}}\prod_{c\in C}P\left(\frac{x}{(\sqrt{2})^N}+\bar{x}\sum_j^N\frac{(-1)^{c_j}}{(\sqrt{2})^j}\right)\,,
\end{equation}
which involves a product of all $2^N$ copies. The normalisation is given by
\begin{equation}
\mathcal{N}=\int \prod_{c\in C}P\left(\frac{x}{(\sqrt{2})^N}+\bar{x}\sum_j^N\frac{(-1)^{c_j}}{(\sqrt{2})^j}\right)dx\,.
\end{equation}
For universality, set $\bar{x}=0$ so that the distilled distribution simplifies to that of the main text,
\begin{equation}
Q_N(x)=\frac{1}{\mathcal{N}}P\left(\frac{x}{(\sqrt{2})^N}\right)^{2^N}\,,
\end{equation}
and this covers the distillation portion of Fig.~\ref{Operational}. The displacement portion can be modelled by the action of the displacement operator $D(\alpha)$, standard in quantum optics, and results in the distribution $Q_N(x-a)$, where the relation between $\alpha$ and $a$ depends on the choice of quadrature probability distribution.

Filtration corresponds to splitting of the distilled, displaced distribution $Q_N(x-a)$ on a variable beamsplitter of transmissivity $T$ alongside the harmonic oscillator ground state. Again using variables labelled $x_1$ and $x_2$, the filtration (including the universal conditioning on $x_2=0$) results in the output distribution
\begin{equation}
Q_N(\sqrt{T}x-a)\frac{e^{-(\sqrt{1-T})x^2}}{\sqrt{\pi}}\,.
\end{equation}
The variance of this output distribution is minimised against $T$, and noise less than that of the oscillator ground state indicates the presence of distillable squeezing.

\section{Distillation Efficiency and Number of Copies}

\begin{figure}
\centering
\includegraphics[width=0.5\columnwidth]{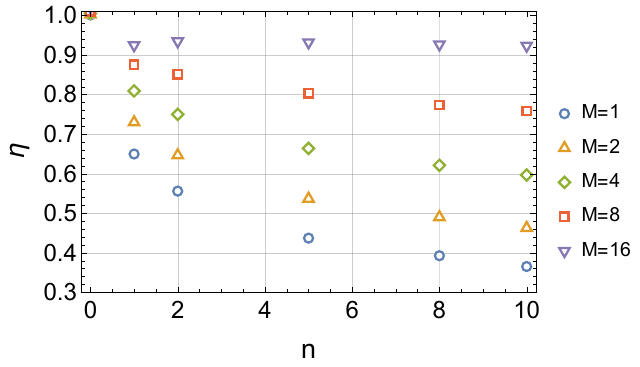}
\caption{The efficiency calculated as the ratio of the asymptotic distillable squeezing to the actual distillable squeezing (so as to give a number between zero and one), for the pure states.}
\label{Eta}
\end{figure}
We define the efficiency $\eta=\frac{\text{Var}(x)_\text{Asym}}{\text{Var}(x)}$, where $\text{Var}(x)_\text{Asym}$ is the asymptotically available squeezing. The efficiencies for the pure states are presented in Fig.~\ref{Eta}. As $n$ increases, the gain in efficiency by increasing the number of copies increases dramatically. That is, for states with greater nonclassical sub-Planck structure there is more to be gained by performing the distillation with a greater number of copies. 

\section{Preservation of Relative Concavity}

\begin{figure}
\centering
\includegraphics[width=0.5\columnwidth]{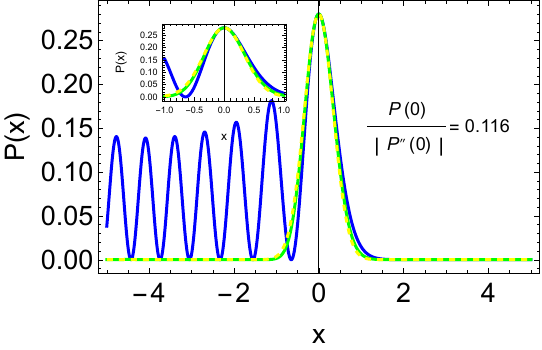}
\caption{Illustration of the preservation of relative concavity for the Fock state $\ket{10}$. The probability distributions for $\ket{10}$ (blue) with global maximum displaced to the origin, displaced $\ket{10}$ after distillation with $M=16$ copies (green), and the asymptotic Gaussian distribution (yellow dashed), where we have rescaled the values of the distribution so that the the distilled and asymptotic distributions have the same global maximum at $x=0$. All three distributions share the same relative concavity, calculated before the rescaling.}
\label{RelConc}
\end{figure}

The method of universal distillable squeezing exposes the nonclassicality of the sub-Planck structure by promoting the substructure around a global maximum of a quadrature probability distribution to a full Gaussian structure, whilst preserving the relative concavity around this maximum. This is illustrated for the pure Fock state $\ket{10}$ in Fig.~\ref{RelConc}.

\section{Sub-Planck Structure in Cat States, GKP States, and Cubic Phase States}

\begin{figure}
\centering
\includegraphics[width=0.49\columnwidth]{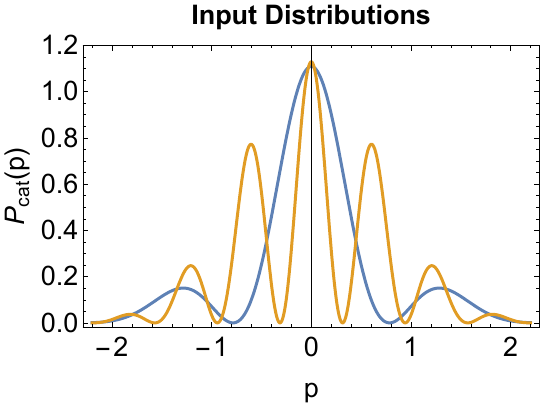}
\includegraphics[width=0.49\columnwidth]{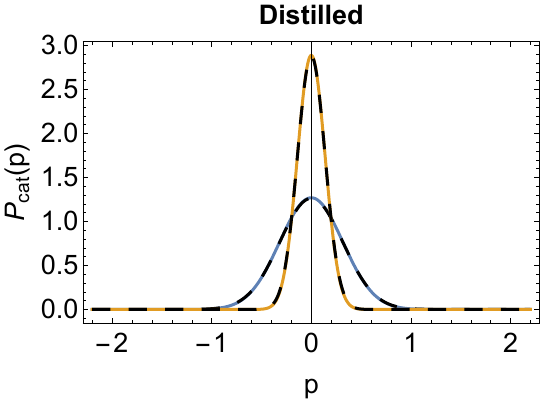}
\caption{The input momentum probability distribution for the cat state with $\alpha=2$ (blue) and $\alpha=5$ (yellow) and the distilled versions with $M=16$ copies (solid colours) and their asymptotic distributions (dashed black). }%Right: The asymptotic distillable squeezing evaluated at the central global maximum, which increases symmetrically with $\alpha$.}
\label{cat}
\end{figure}
Sub-Planck structures are almost universally present in non-Gaussian quantum states. Here we show the effectiveness of the universal distillation procedure tested in the main text to pure cat states, pure GKP states, and pure cubic phase states. These states are now phase sensitive, so the choice of phase space variable matters. Additionally, we show that the failure of the universal method for a small class of cubic phase states can be overcome by dropping the universal property and applying a non-universal distillation of squeezing to unveil nonclassical sub-Planck structures that are even more obscured than those associated with the relative concavity of the global maximum.

The cat states are composed of superpositions of coherent states $\ket{\alpha}$, where we select $\alpha\in\mathbb{R}$ without loss of generality. For cat states of the form $\ket{\psi_\text{cat}}\propto\ket{\alpha}+\ket{-\alpha}$, the momentum probability distribution can be expressed
\begin{equation}
P_\text{cat}(p)=\frac{e^{-(p+i\alpha)^2}(1+e^{2ip\alpha})^2}{2\sqrt{\pi}(1+e^{\alpha^2})}\,.
\end{equation}
This distribution has a global maximum at $p=0$ for all $\alpha$. As $\alpha$ increases, the central peak narrows while further smaller peaks emerge on either side. An example distribution and the asymptotic distillable squeezing are displayed in Fig.~\ref{cat}. The greater the visibility of the interference ripples, the greater the nonclassicality of the sub-Planck structures evinced by the distillable squeezing.

\begin{figure}
\centering
\includegraphics[width=0.5\columnwidth]{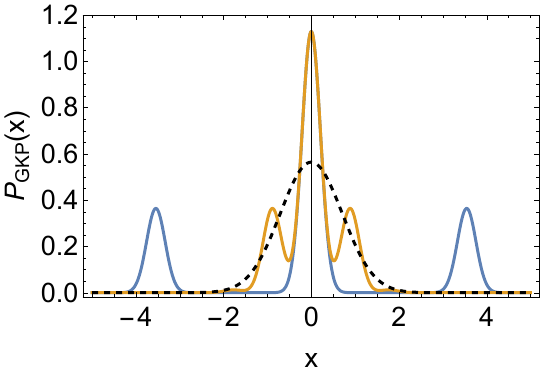}
\caption{The position probability distribution for the approximate GKP states with $\Delta=\frac{3}{10}$ and $S=3$, and with $\sqrt{\pi}$ spacing (blue), and $\frac{\sqrt{\pi}}{4}$ spacing (yellow). The ground state is the black dashed curve. The reduced spacing broadens the central peak resulting in an increased variance above the squeezing threshold, even when discarding outer peaks. The relative concavity is almost perfectly maintained however, allowing distillable squeezing to almost perfectly recover the original squeezing of the central Gaussian squeezed element of the superposition.}
\label{ProbGKP}
\end{figure}

The ideal GKP state is composed of superpositions of position eigenstates, and can be approximated by a superposition of squeezed states under a Gaussian envelope. The resulting position probability distribution can be approximated as~\cite{gottesman_encoding_2001}
\begin{equation}
P_\text{GKP}=\frac{2}{\sqrt{\pi}}\sum_{s=-S}^{S}e^{-4\pi\Delta^2s^2}e^{-\frac{(x-2s\sqrt{\pi})^2}{\Delta^2}}\,,
\end{equation}
where $S$ defines the number of peaks to the left and right of the central peak (ideally infinitely large), and $\Delta$ defines the Gaussian suppression of peaks far from the origin. Since the peaks are well-separated in this approximation, the asymptotic distillable squeezing is equal to the squeezing of of the central element of the superposition at $s=0$, which has variance $\frac{\Delta^2}{2}$, verified numerically. This is well approximated by cutting out the side peaks by integrating over the narrow region which includes only the central peak. However, when the spacing of $\sqrt{\pi}$ is reduced, as under conditions of pure loss, the central peak eventually becomes broadened by the overlap with the side peaks. In this realistic case, the distillable squeezing cannot be estimated by merely examining the width of the central peak. However the distillable squeezing still recovers the original squeezing of the central peak in the superposition, as shown in Fig.~\ref{ProbGKP}. For the numbers used here, the central element of the superposition has variance $0.125$. When the spacing is reduced, this variance increases to approximately 2.35, however the distillable squeezing still recovers a variance of $0.128$, almost the original value and far below the variance estimated by the broadened central peak.

The cubic phase state results from the application of the cubic phase gate to the oscillator ground state $\ket{0}$. The state can be expressed as $\ket{\psi_{cub}}=e^{i\frac{\gamma}{3}q^3}\ket{0}$, where $q$ is the position operator. The probability distribution in momentum is
\begin{equation}
P_\text{cub}(p)=\left(\frac{\sqrt{2}\pi^{\frac14}}{\gamma^{\frac13}}e^{\frac{1-\gamma p}{12\gamma^2}}\text{Ai}\left(\frac{1-4\gamma p}{4\gamma^{\frac43}}\right)\right)^2\,.
\end{equation}
The major peak of this distribution becomes broader as $\gamma$ increases, rather than narrower, so that relative concavity cannot be converted to distillable squeezing. This limitation is overcome by non-universal distillation (see also SM Operational Derivation of Distillable Squeezing). 
\begin{figure}
\centering
\includegraphics[width=0.5\columnwidth]{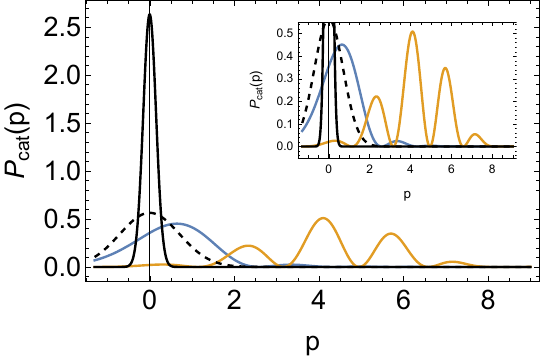}
\caption{The momentum probability distribution for the cubic phase state with $\gamma=1$ (blue), and after a single non-universal distillation layer with $\bar{x}=5$ (yellow). The non-universal distillation layer prepares interferences fringes with increased concavity around the global maximum, whose squeezing may then be universally distilled (black) to reveal the nonclassical sub-Planck structure in the form of distillable squeezing, with the asymptotic value $\text{Var}(x)=0.152$.}
\label{ProbCub}
\end{figure}

With a single distillation step involving two copies of the distribution, conditioned on a measurement outcome $\bar{x}$ that is not the global maximum of the distribution, the output distribution possesses a narrower peak around the global maximum, as in Fig.~\ref{ProbCub}. This increased concavity can then be submitted to the universal method described in the main text in order to see distillable squeezing and therefore unveil nonclassicality that would have been obscured from the a pure application of the universal method. We emphasise that the non-universal step also cannot prepare nonclassical states from classical ones, and is therefore faithful, and can also be carried out on a measured on deduced distribution, which does not require physical implementation.

\end{document}